\documentclass[a4paper,11pt]{article}
\usepackage{multirow}
\usepackage{amsfonts}
\usepackage{adjustbox}
\usepackage{amsmath}
\usepackage{graphicx}
\usepackage{pict2e}
\usepackage{epsfig}
\usepackage[
    font      = footnotesize,
    labelfont = bf
]{caption}
\usepackage[
    font      = footnotesize,
    labelfont = bf
]{subcaption}

\usepackage[colorlinks=true]{hyperref}

\begin{document}
\newcommand{\ds}{\displaystyle}
\newcommand{\be}{\begin{equation}}
\newcommand{\ee}{\end{equation}}
\newcommand{\ba}{\begin{array}}
\newcommand{\ea}{\end{array}}
\newcommand{\bea}{\begin{eqnarray}}
\newcommand{\eea}{\end{eqnarray}}

\def\R{{\mathbb R}}
\def\C{{\mathbb C}}
\newcommand{\bi}{\begin{itemize}}
\newcommand{\ei}{\end{itemize}}
\newcommand{\sn}{\mbox{sn}}
\newcommand{\cn}{\mbox{cn}}
\newcommand{\dn}{\mbox{dn}}
\newcommand{\x}{{\ensuremath{\times}}}
\newcommand{\bb}[1]{\makebox[16pt]{{\bf#1}}}
\newtheorem{theorem}{Theorem}
\newtheorem{definition}{Definition}
\newtheorem{lemma}{Lemma}
\newtheorem{comment}{Comment}
\newtheorem{corollary}{Corollary}
\newtheorem{example}{Example}
\newtheorem{examples}{Examples}
 \newtheorem{conjecture}{Conjecture}
\title{Separation equations for 2D superintegrable systems on constant curvature spaces}
\author{{\bf M.A. Escobar-Ruiz}\\
{\sl Instituto de Ciencias Nucleares, UNAM}\\
{ \sl Apartado Postal 70-543, 04510 Mexico D.F. MEXICO}\\
{\sl and School of Mathematics, University of Minnesota}\\
{\sl mauricio.escobar@nucleares.unam.mx}\\
{\bf Ernest G. Kalnins}\\
{\sl Department of Mathematics,
University
of Waikato,}\\
{\sl Hamilton, New Zealand}\\
{\sl   and }  {\bf Willard Miller, Jr.}\\
{\sl School of Mathematics, University of Minnesota,}\\
{\sl Minneapolis, Minnesota,
55455, U.S.A.}\\
{\sl miller@ima.umn.edu}}
\date{\today}
\date{\today}
\maketitle

\begin{abstract}
Second-order conformal quantum superintegrable systems in $2$ dimensions are Laplace equations on a manifold with an added scalar potential 
and $3$ independent 
2nd order conformal
symmetry operators. They encode all the information
about 2D Helmholtz or time-independent Schr\"odinger   superintegrable systems in an efficient manner:  
Each of these systems admits a quadratic symmetry algebra (not usually a Lie algebra) and is multiseparable. We study the separation equations for the systems 
as a family rather than separate cases.
We show that the separation equations comprise all of the various types of hypergeometric and Heun equations in full generality. In particular, 
they yield all of the 1D 
Schr\"odinger exactly solvable (ES) and quasi-exactly solvable (QES) systems related to the Heun operator. 
We focus on complex constant curvature spaces and show explicitly that there are 8 pairs of Laplace separation types and these types
account for all separable coordinates  on the 20 flat space and 9 2-sphere Helmholtz superintegrable systems, including those for the constant potential case. 
The different systems are related by St\"ackel transforms, by the symmetry algebras
and by B\"ocher contractions of the conformal algebra $so(4,\C)$ to itself, which enables all  systems to be derived from a single one:
the generic potential on the 
complex 2-sphere. This approach facilitates a unified view of special function theory, incorporating hypergeometric and Heun functions
in full generality.
\end{abstract}

\section{Introduction}

We show how Laplace superintegrable systems theory unifies and simplifies the theory of the special functions of mathematical physics, 
hypergeometric and Heun equations, and exactly solvable and quasi-exactly solvable systems, through the study of the family of separation equations for these systems. Quantum superintegrable systems are those with maximal symmetry,
which permits their explicit solution. We consider here one of the simplest classes of such systems: 2nd order superintegrable systems in 2 
complex (or real) variables. 
This  is an integrable Hamiltonian system on an $2$-dimensional Riemannian/pseudo-Riemannian manifold 
with potential: 
\[ H=\Delta_2+V,\]   that admits $3$ 
algebraically independent  2nd order partial differential operators $L_1,L_2,H$ commuting with $H$, the maximum  possible, \cite{MPW2013},
\[ [H,L_j]=0,\quad   \ j=1,2.\] 
Here $[A,B]=AB-BA$ is the operator commutator. 
Superintegrability captures the properties of 
quantum Hamiltonian systems that allow the Schr\"odinger eigenvalue problem $H\Psi=E\Psi$ to be solved exactly, analytically and algebraically. 
The 2nd order 2D systems have been classified. 
There are 37  nondegenerate (3 linear parameter potential) systems, on a variety of manifolds, and 21 degenerate (1 parameter potential) systems. 
The generating symmetries of every such  system form a quadratic algebra \cite{MPW2013}.
For simplicity, here we consider only the systems on flat space and the 2-sphere (listed in Appendix A); the treatment for other spaces is  similar.

 The fundamental system is 
 quantum S9, the generic potential on the complex 2-sphere, $s_1^2+s_2^2+s_3^2=1$.
This system has generating symmetries
\begin{align} H=&J_1^2+J_2^2+J_3^2+\frac{a_1}{s_1^2}+\frac{a_2}{s_2^2}+\frac{a_3}{s_3^2},\\
        L_1= &J_1^2+\frac{a_3 s_2^2}{s_3^2}+\frac{a_2 s_3^2}{s_2^2},\
        L_2= J_2^2+\frac{a_1 s_3^2}{s_1^2}+\frac{a_3s_1^2}{s_3^2}, \label{S9Operators}\end{align}
        where $J_3=s_1\partial_{s_2}-s_2\partial_{s_1}$ and cyclic permutations.
The algebra is given by
\bea \ba{rl} [L_i,R]=&\!\!\!\!4\{L_i,L_k\}-4\{L_i,L_j\}- (8+16a_j)L_j + (8+16a_k)L_k+ 8(a_j-a_k),\\
 R^2=&\!\!\!\!\frac83\{L_1,L_2,L_3\} -(16a_1+12)L_1^2 -(16a_2+12)L_2^2  -(16a_3+12)L_3^2\\
&\!\!\!\!+\frac{52}{3}(\{L_1,L_2\}+\{L_2,L_3\}+\{L_3,L_1\})+ \frac13(16+176a_1)L_1\\
&\!\!\!\!+\frac13(16+176a_2)L_2 + \frac13(16+176a_3)L_3 +\frac{32}{3}(a_1+a_2+a_3)\\ 
&\!\!\!\!+48(a_1a_2+a_2a_3+a_3a_1)+64a_1a_2a_3.\ea \label{S9Structure}\eea
Here, $R=[L_1,L_2]$ and  $\{i,j,k\}$ is a cyclic permutation of $\{1,2,3\}$, and to simplify the structure equations we set  $L_3=H-L_1-L_2-a_1-a_2-a_3$.

The Higgs oscillator (Quantum  $S3$) is the fundamental degenerate system. It is the same as $S9$ with $a_1=a_2=0$, $a_3=a$ but admits additional symmetry. The symmetry algebra is generated by 
\bea\label{S3Operators} X=J_3, \qquad L_1=J_1^2 + \frac{a s_2^2}{s_3^2}, \qquad 
L_2= \frac12(J_1J_2+J_2J_1)-\frac{a s_1s_2}{s_3^2}\nonumber .\eea
The structure relations for the algebra are given by $R=[L_1,L_2]$ and
\bea \label{S3Structure}
\ba{l}
\left[ L_1,X \right]=2L_2,\qquad [L_2,X]= -X^2-2L_1+H-a,\\
\left[ L_1,L_2 \right]=-(L_1X+XL_1)-(\frac12+2a)X,\\
0=\{L_1, X^2\}+2L_1^2+2L_2^2-2L_1H+\frac{5+4a}{2}X^2-2aL_1-a.\ea\eea

All these systems can be treated more conveniently as Laplace equations, \cite{BocherCon1,BocherCon2}. Since every 2D manifold is conformally flat, there always exist ``Cartesian-like'' coordinates 
$x,y$ such that $H=\frac{1}{\lambda(x,y)}(\partial_x^2+\partial_y^2)+V(x,y)$. Thus the Helmholtz equation $H\Psi=E\Psi$ on some conformally flat space is 
equivalent to the Laplace equation (with potential) 
\be\label{tildeV} (\partial_x^2+\partial_y^2 +{\tilde V}(x,y))\Psi=0\ee
on flat space,  where ${\tilde V}=\lambda(V-E)$, so the eigenvalue $E$ has been incorporated as a parameter in the new potential.

More generally, we consider Laplace systems of the form
  \be\label{Laplace} H\,\Psi({\bf x}) \ \equiv \ (\,\Delta_2+V({\bf
x})\,)\,\Psi({\bf x}) \ =\ 0\ .\ee
Here $\Delta_2 $ is the Laplace-Beltrami operator on a real or complex
$2D$ Riemannian or pseudo-Riemannian manifold and $V$ is
a  non-zero scalar potential. All variables can be complex, except when we impose constraints such as square integrability.
A conformal symmetry of  equation (\ref{Laplace}) is a partial
differential operator  $L$ in the variables ${\bf x}=(x_1,\,x_2)$ such
that $[ L, H]\equiv LH-HL=R_{ L} H$ for some differential operator $R_{L}$.
A conformal symmetry maps any solution $\Psi$ of (\ref{Laplace}) to
another solution. Two conformal symmetries ${ L}, { L}'$ are
identified if $L=L'+S\,H$ for some differential operator $S$, since they
agree on the solution space of (\ref{Laplace}). (For brevity we will say
that $L=L', \mod (H)$
and that $L$ is a conformal symmetry if $[L,H]=0,\mod(H)$.)
The system is {\it conformally superintegrable} if there exist three
algebraically independent conformal symmetries,
${ L}_1,\,{L}_{2},\, L_3$ with ${L}_3={ H}$. It is second order
conformally superintegrable if 
$L_2$ can be chosen to be a 2nd order differential operator, and $L_1$
of at most 2nd order. (If the system admits symmetries such that
$L_1,L_2$ can be chosen as 1st order, we
say it is 1st order conformally superintegrable). 

An important fact is that the mapping of a Helmholtz superintegrable system $H\Psi=E\Psi$
to the Laplace equation (\ref{tildeV}) preserves superintegrability, i. e., system 
(\ref{tildeV}) is conformally superintegrable, \cite{BocherCon1,BocherCon2}. Further, 
Suppose we have a second order conformal  superintegrable system 
\[  H=\partial_{xx}+\partial_{yy}+V(x,y)=0,\quad   H= H_0+V\]
where $V(x,y)= W(x,y)-E\, U(x,y)$ for arbitrary parameter $E$. The potential $U$ defines a {\it conformal St\"ackel
transform} to  the  (Helmholtz) system  \[{\tilde {  H}}\Psi=E\Psi,\quad 
{\tilde { H}}=\frac{1}{{ U}}(\partial_{xx}+\partial_{yy})+{\tilde V}
\]
 where
  ${\tilde V}=\frac{W}{U}$. and this Helmholtz system is superintegrable, \cite{BocherCon1,BocherCon2}.
The Laplace system on flat space has metric $ds^2=dx^2+dy^2$, and volume element $dx\ dy$,
whereas the St\"ackel transformed system has metric $ds^2=U(dx^2+dy^2)$ and volume element $ U\ dx\ dy$.
There is a similar definition of St\"ackel transforms of  Helmholtz superintegrable systems $H\Psi=E\Psi$  which take superintegrable 
systems to superintegrable systems, essentially preserving the quadratic algebra structure.
Thus  any second order conformal Laplace superintegrable system admitting a nonconstant potential $U$ can be 
St\"ackel transformed to a Helmholtz
superintegrable system.

A crucial observation here is that each equivalence class of Laplace superintegrable systems is multiseparable, \cite{MPW2013,KKM20042,KKM20061,EM2016},
and that the separable coordinates 
are invariant under St\"ackel tranforms. Thus all Helmholtz systems in an equivalence class are separable in {\it exactly the same coordinates}. 
The separable coordinates may look different when subjected to a St\"ackel transform but they are unchanged.
 The  equivalence classes of Laplace nondegenerate and degenerate superintegrable systems with functionally independent generators are listed in tables 
\ref{Tab1} and \ref{Tab2}. For each Laplace system (given in Cartesian coordinates $x,y$) we list 
the possible separable orthogonal coordinates and the constant curvature Helmholtz systems that arise from it via St\"ackel transforms. 
To identify the relevant St\"ackel tranforms we use the notation ${\bf a}=(a_1,a_2,a_3,a_4)$ to represent the nondegenerate Laplace 
potential $V=\sum_i a_i V^{(i)}$
and ${\bf a}=(a_1,a_2)$ to represent the degenerate Laplace potential $V= a_1 V^{(1)}+ a_2 V^{(2)}$.
For a fixed choice of parameters ${\bf b}=(b_1,b_2,b_3,b_4)$, ${\bf b}=(b_1,b_2)$, respectively, each St\"ackel transform is 
described by $\frac{1}{V({\bf b})}\left(\partial_x^2+\partial_y^2
+\frac{V({\bf a})}{V({\bf b})}\right)=0$. In the tables we list the possible transforms that correspond to constant curvature Helmholtz systems.

{\small
{
\begin{table}[tbhp]
\begin{center}
\resizebox{\textwidth}{!}{%
\begin{tabular}{|l||l| }
\hline
   &   \\
\multirow{2}{*}{\qquad\qquad Laplace Systems}  &    Non-degenerate potentials: \quad $\partial_x^2+\partial_y^2+V(x,\,y)$  \\ [1ex]
\hline
\hline
  &   \\
$ [1111] $ & $V(x,y)=
\frac{a_1}{x^2}+\frac{a_2}{y^2}+\frac{4\,a_3}{(x^2+y^2-1)^2}-\frac{4\,a_4}{(x^2+y^2+1)^2}$\\ [1.1ex]
\quad Laplace separable coordinates:&\\
 &$a)$\ spherical:\quad  $x=\frac{\sin(\theta)\,\cos(\phi)}{1+\cos(\theta)},\ y=\frac{\sin(\theta)\,\sin(\phi)}{1+\cos(\theta)}$\\ [1.1ex]
& $b)$\ ellipsoidal:\quad $x^2=\frac{(cu-1)(cv-1)}{(1-c)(1+\sqrt{cuv})^2},\ y=\frac{c(u-1)(v-1)}{(c-1)(1+\sqrt{cuv})^2},\ c\ne 0,1$,\\ [1.1ex]
\quad Helmholtz superintegrable systems:& $S9$: $(1,0,0,0),(0,1,0,0,),(0,0,1,0),(0,0,0,1)$\\ [1.1ex]
 & $S8$: $(1,1,1,1),(0,1,0,1),(1,0,1,0),(0,1,1,0),(1,0,0,1)$\\ [1.1ex]
& $S7$: $(1,1,0,0),(0,0,1,1)$
\\ [1.8ex]\hline &\\
$ [211]  $ & $V(x,y)= \frac{a_1}{x^2} + \frac{a_2}{y^2} - a_3\,(x^2+y^2) + a_4  $\\[1.1ex]
\quad Laplace separable coordinates:&\\
&$c)$\ Cartesian:\quad $x,y$,\\ [1.1ex]
& $d)$\ polar:\quad $x=r\cos(\theta),\ y=r\sin(\theta)$,\\ [1.1ex]
&$e)$\  elliptic: $x=c\sqrt{(u-1)(v-1)},\ y=c\sqrt{-uv}$,\\ [1.1ex]
\quad Helmholtz superintegrable systems:& $S4$: $(1,1,0,0)$\\[1.1ex]
& $S2$: $(1,0,0,0),(0,1,0,0)$\\[1.1ex]
& $E1$: $(0,0,0,1)$\\[1.1ex]
& $E16$: $(0,0,1,0)$
\\  [1.8ex] \hline&\\
$ [22]    $ & $V(x,y)=
\frac{a_1}{(x+i\,y)^2}+\frac{a_2\,(x-i\,y)}{(x+i\,y)^3}+a_3-a_4\,(x^2+y^2) $\\ [1.1ex]
\quad Laplace separable coordinates:&\\
&$d)$\ polar\\[1.1ex]
&$f)$\ hyperbolic: \quad $x=\frac{u^2+v^2+u^2v^2}{2uv},\ y=\frac{i(u^2+v^2-u^2v^2)}{2uv}$\\[1.1ex]
\quad Helmholtz superintegrable systems:& $E7$: $(1,0,a_3,0)$\\ [1.1ex]
& $E8$: $(0,0,1,0)$\\ [1.1ex]
& $E17$: $(0,0,0,1)$\\ [1.1ex]
& $E19$: $(0,1,0,a_4)$
\\  [1.8ex]
\hline&\\
$ [31]    $ & $V(x,y)= a_1-a_2\,x+a_3\,(4\,x^2+y^2)+\frac{a_4}{y^2} $\\ [1.1ex]
\quad Laplace separable coordinates:&\\
&$c)$\ Cartesian \\ [1.1ex]
&$g)$\ parabolic: \quad $x=\xi^2-\eta^2,\ y=2\xi\eta$,\\ [1.1ex]
\quad Helmholtz superintegrable systems: & $S1$ $(0,0,0,1)$\\ [1.1ex]
& $E2$: $(1,0,0,0)$
\\  [1.8ex]
\hline&\\
$ [4]      $ & $V(x,y)= a_1-a_2\,(x+i\,y)
+a_3\,(3(x+i\,y)^2+2(x-i\,y))-a_4\,(4(x^2+y^2)+2(x+i\,y)^3) $\\ [1.1ex]
\quad Laplace separable coordinates:&\\
&$h)$\ semi-hyperbolic:\quad $x=-(w-u)^2+i(w+u),\ y=-i(w-u)^2+(w+u)$,\\[1.1ex]
\quad Helmholtz superintegrable systems:& $E9$: $(0,1,0,0)$\\ [1.1ex]
& $E10$: $(1,a_2,0,0)$
\\  [1.8ex]
\hline&\\
$ [0]      $ & $V(x,y)= a_1-(a_2\,x+a_3\,y)+a_4\,(x^2+y^2)   $ \\ [1.1ex]
\quad Laplace separable coordinates:&\\
&$c)$\  Cartesian\\ [1.1ex]
\quad Helmholtz superintegrable systems:&  $E3'$: $(1,0,0,0)$\\ [1.1ex]
& $E11$: $(a_1,1,\pm i,0)$\\ [1.1ex] 
& $E20$: $(\frac{a_2^2+a_3^2}{4},a_2,a_3,1)$
\\  [1.5ex]
\hline
\end{tabular}}
\caption{Four-parameter potentials for Laplace systems. Each
of the Helmholtz nondegenerate constant curvature superintegrable
(i.e., 3-parameter) eigenvalue systems is St\"ackel equivalent
to exactly one of these systems.}
\label{Tab1}
\end{center}
\end{table}
}

\begin{table}[tbhp]
\small
\setlength{\tabcolsep}{2pt}
\begin{tabular}{|l||l| }
\hline
   &   \\
\multirow{2}{*}{Laplace Systems}  &    Degenerate potentials \quad $\partial_x^2+\partial_y^2+V(x,\,y)$  \\ [1ex]
  \hline
\hline
  &   \\
$ A $ & $V(x,y)= \frac{4\,a_3}{(x^2+y^2-1)^2}-\frac{4\,a_4}{(x^2+y^2+1)^2} $\\[1.1ex]
\quad Laplace separable coordinates: & $a)$\ spherical\quad $b)$\ ellipsoidal\\[1.1ex]
& $i)$\ horospherical: \quad $x=\frac12\frac{u^2+v^2-1}{u-iv},\ y=-\frac{i}{2}\frac{u^2+v^2+1}{u-iv}$,\\[1.1ex]
 & $j)$\ degenerate elliptic: \quad $x=\frac{u^2+v^2}{4uv}+\frac{uv}{u^2v^2+1}$,\\[1.1ex] 
&\qquad \qquad \qquad\qquad\quad\ \  $y=-\frac{i(u^2+v^2)}{4uv}+\frac{iuv}{u^2v^2+1}$\\[1.1ex]
\quad Helmholtz superintegrable systems:& $S3$: $(1,0),(0,1)$\qquad
 $S6$: $(1,1)$
\\ [1.5ex]
\hline&\\
$ B  $ & $V(x,y)= \frac{a_1}{x^2}+a_4 $\\[1.1ex]
\quad Laplace separable coordinates:& $c)$\ Cartesian\quad $d)$\ polar\\[1.1ex]
& $g)$\ parabolic\quad  $e)$\ elliptic\\[1.1ex]
\quad Helmholtz superintegrable systems:& $S5$: $(1,0)$\qquad
 $E6$: $(0,1)$
\\  [1.5ex]
\hline&\\
$ C    $ & $V(x,y)= a_3-a_4\,(x^2+y^2) $\\[1.1ex]
\quad Laplace separable coordinates:& $c)$\ Cartesian\quad $d)$\ polar\\[1.1ex]
&$f)$\ hyperbolic\quad $e)$\ elliptic\\[1.1ex]
\quad Helmholtz superintegrable systems:& $E3$: $(1,0)$\qquad 
$E18$: $(0,1)$
\\  [1.5ex] \hline&\\
$ D    $ & $V(x,y)= a_1-a_2\,x $\\[1.1ex]
\quad Laplace separable coordinates:& $c)$\ Cartesian\quad $g)$\ parabolic\\[1.1ex]
\quad Helmholtz superintegrable systems:& $E5$: $(1,0)$
\\  [1.5ex]  \hline&\\
$ E      $ & $V(x,y)= \frac{a_1}{(x+i\,y)^2}+a_3  $\\[1.1ex]
\quad Laplace separable coordinates:& $d)$\ polar\quad $f)$\ hyperbolic\\[1.1ex]
\quad Helmholtz superintegrable systems:& $E12$: $(a_1,a_3)$, $a_1a_3\ne 0$\qquad
$E14$: $(0,1)$
\\  [1.5ex] \hline&\\
$ F     $ & $V(x,y)= a_1-a_2\,(x+i\,y)  $\\[1.1ex]
\quad Laplace separable coordinates:{\hfill }&$c)$\ Cartesian\quad  $h)$\ semi-hyperbolic\\[1.1ex]
\quad Helmholtz superintegrable systems:& $E4$: $(1,0)$,\qquad
 $E13$: $(a_1,a_2)$, $a_2\ne 0$
\\  [2.0ex]
\hline
\end{tabular}
\caption{Two-parameter degenerate potentials for Laplace systems.}
\label{Tab2}
\end{table}
}

\section{B\^ocher contractions}
\begin{center}
\begin{figure} [tbhp]
\includegraphics[width=12.0cm]{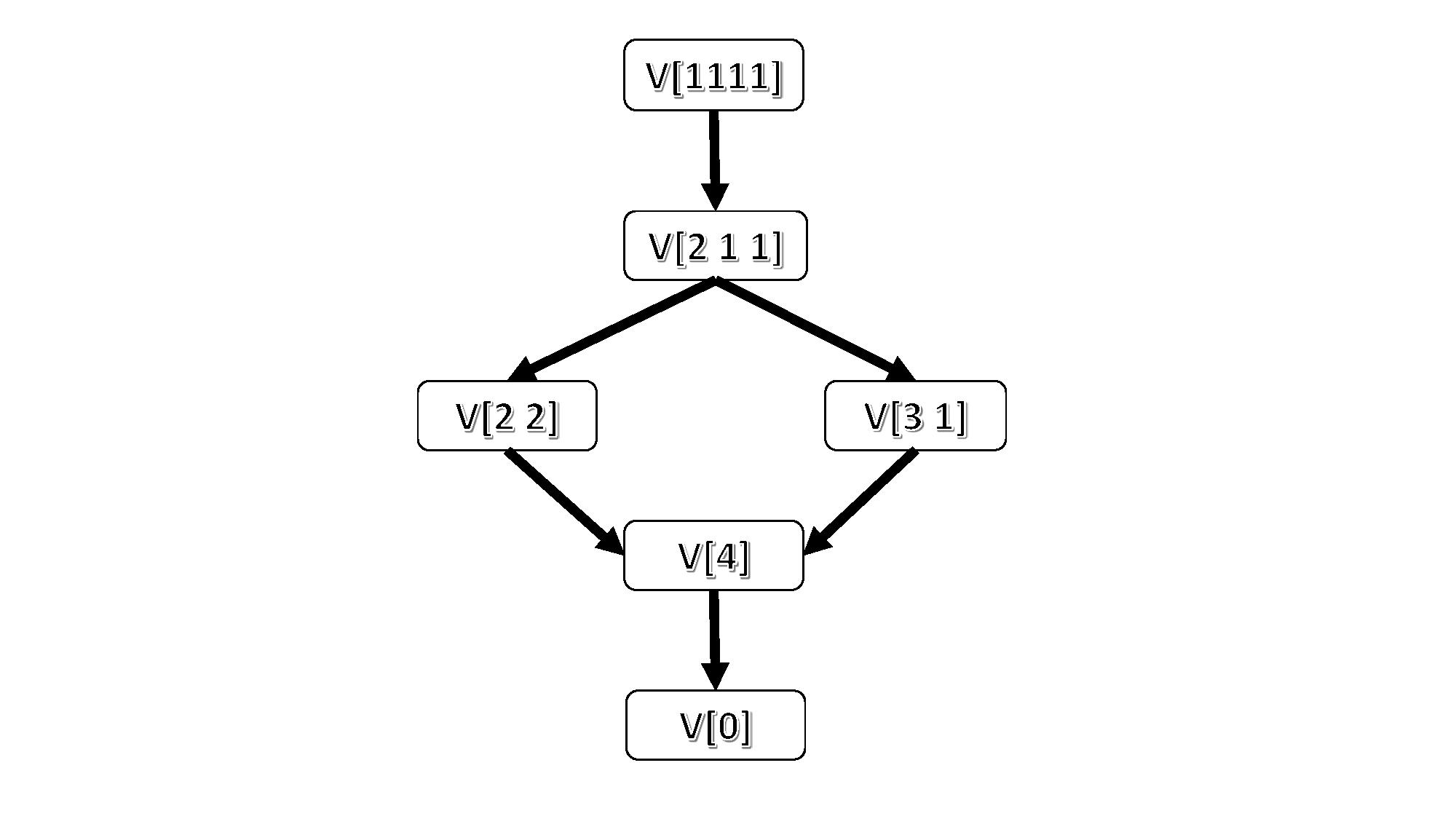}
\caption{Contractions of nondegenerate Laplace systems }
\label{var2}
\end{figure}
\end{center}

\begin{center}
\begin{figure}[tbhp]
\includegraphics[width=12.0cm]{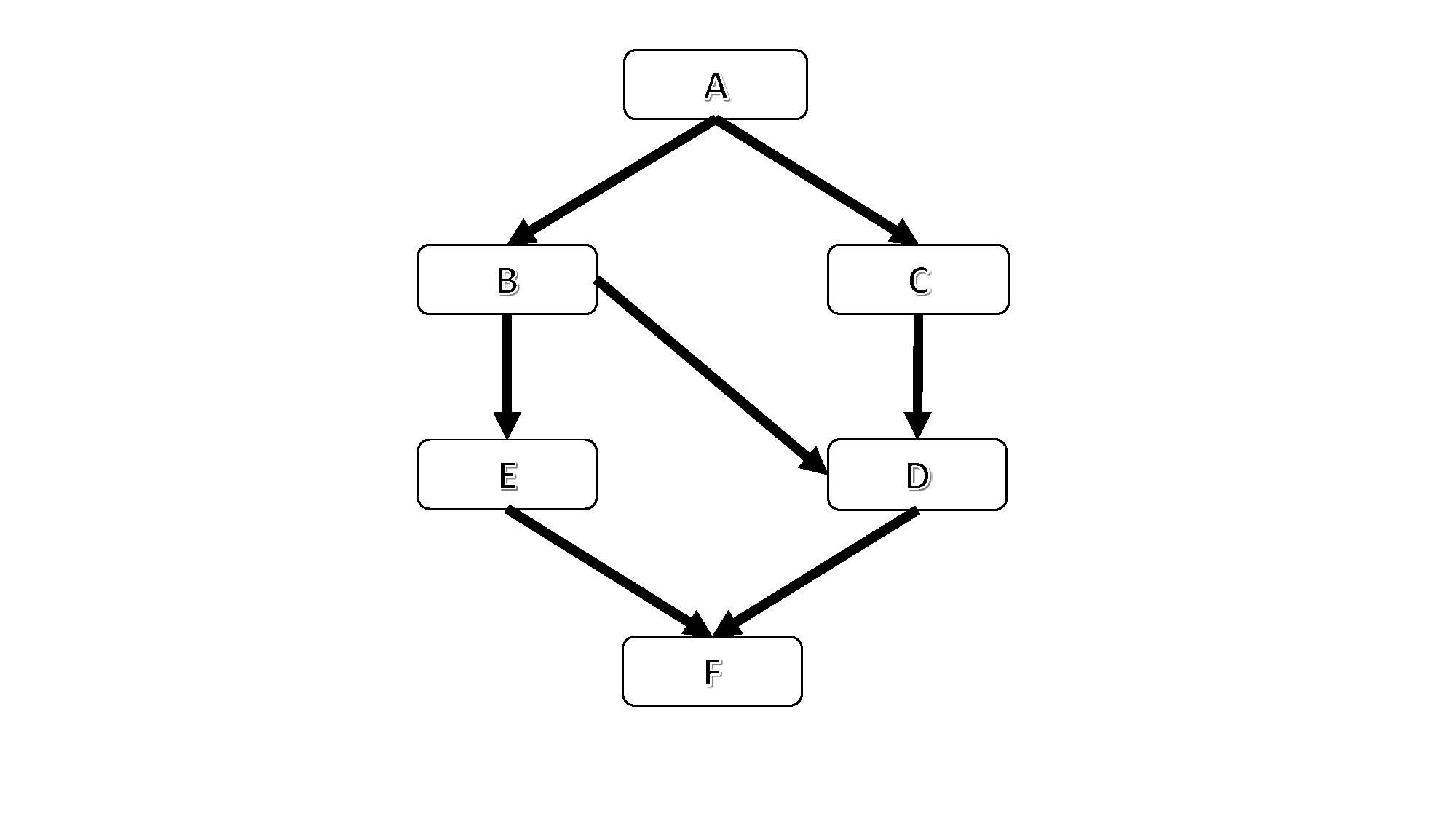}
\caption{Contractions of degenerate Laplace systems }
\label{var3}
\end{figure}
\end{center}

All Laplace conformally superintegrable systems can be obtained as limits of the basic system $[1111]$, \cite{BocherCon1,BocherCon2}.
The conformal symmetry algebra of the underlying flat space free Laplace equation is $so(4, {\C})$, and these limits are described by 
Lie algebra contractions of this conformal algebra to itself, which can be classified. We call these B\"ocher contractions since they 
are motivated by ideas of B\"ocher, \cite{Bocher}, who used similar limits to construct separable coordinates of free  Laplace, 
wave and Helmholtz equations from basic cyclidic coordinates. (A major difference here is that our systems include potentials.) 
There are 4 basic B\"ocher contractions of 2D Laplace  systems and each one when applied to a Laplace system yields another Laplace 
superintegrable system, \cite{BocherCon1,BocherCon2}. These in turn induce contractions of the Helmholtz systems in each equivalence 
class to Helmholtz systems 
in other classes, over 200 contractions in all. However, we can summarize the basic results for Laplace systems in 
Figures  \ref{var2} and  \ref{var3}. A system can be obtained 
from another superintegrable system via contraction provided it is connected to the other system by 
directed arrows. All systems follow from $[1111]$ for nondegenerate potentials and from $A$ for degenerate potentials, 
and $A$ is a restriction of $[1111]$ with increased symmetry.

\section{St\"ackel transforms}

A St\"ackel transformed Helmholtz system from one of the tables 
\ref{Tab1} or \ref{Tab2} will have metric 
\[ds^2=V({\bf b})(dx^2+dy^2).\] 
If the metric is flat, we need to determine coordinates $X(x,y)$,$Y(x,y)$ such that $ds^2=dX^2+dY^2$.  Then separable coordinates  $x(u,v),y(u,v)$ for the Laplace system tranform to separable coordinates $X(x(u,v),y(u.v))=X[u.v]$, $Y(x(u,v),y(u.v))=Y[u.v]$,
for the Helmholtz system. Similarly if the metric corresponds to $S_{2,\C}$, we need to  determine 
coordinates $s_1(x,y)$. $s_2(x,y)$, $s_3(x,y)$ such that $s_1^2+s_2^2+s_3^2=1$ and $ds^2=ds_1^2+ds_2^2+ds_3^2$. 
Then the separable coordinates will be expressed as $ s_1[u,v],s_2[u,v],s_3[u,v]$.
The possible separable coordinate systems $(u,v)$ in flat space and $[u,v]$ on the 2-sphere are listed in Appendix B.

\subsubsection{St\"ackel equivalences}\label{Stackelequiv}
The relevant coordinate transformations are:

 \medskip \noindent
  Flat space - flat space
  \begin{enumerate}
   \item \[ dX^2+dY^2=\frac{dx^2+dy^2}{(x+iy)^2},\
    X=\frac14\frac{x^2+y^2-4}{x+iy},\ Y=\frac{i}{4}\frac{x^2+y^2+4}{x+iy}.\]
   \item \[ dX^2+dY^2=\left(\frac{x-iy}{(x+iy)^3}-a(x^2+y^2)\right)(dx^2+dy^2),\]
   \[ X = -ia(x+iy)^2-\frac{i}{16}\frac{(x^2+y^2)^2+16}{(x+iy)^2},\ Y= -a(x+iy)^2+\frac{1}{16}\frac{(x^2+y^2)^2-16}{(x+iy)^2}.\]
   \item \[ dX^2+dY^2=\left(x^2+y^2\right)(dx^2+dy^2),\
    X=\frac12(x^2-y^2),\ Y=xy.\]
   \item \[ dX^2+dY^2=(x+iy)(dx^2+dy^2),\]
   \[ X=\frac{i}{2}(x+iy)^2-\frac{i}{4}(x-iy),\ Y=\frac{1}{2}(x+iy)^2+\frac{1}{4}(x-iy).\]
   \item \[ dX^2+dY^2=\left(1+a(x+iy)\right)(dx^2+dy^2),\]
   \[ X=-i(x+iy)^2+\frac{ia}{8}(x-iy)-\frac{2i}{a}(x+iy),\  Y=-(x+iy)^2-\frac{a}{8}(x-iy)-\frac{2}{a}(x+iy)-\frac{1}{a^2}.\]
   \item \[ dX^2+dY^2=\left(\frac{1}{(x+iy)^2}+a\right)(dx^2+dy^2),\]
   \[ X=-\frac{i}{4(x+iy)}+\frac{ia}{4}(x+iy)-i(x-iy),\ Y=-\frac{1}{4(x+iy)}+\frac{a}{4}(x+iy)+(x-iy).\]
   \item\[ dX^2+dY^2=(x+iy)(dx^2+dy^2),\]
   \[ X=\frac{i}{2}(x+iy^2)^2-\frac{i}{4}(x-iy),\ Y=\frac12(x+iy)^2+\frac14(x-iy).\]
  \end{enumerate}

\medskip \noindent
 Flat space - sphere
  \begin{enumerate}
\item \[ \frac{4(dx^2+dy^2)}{(x^2+y^2+1)^2}=ds_1^2+ds_2^2+ds_3^2,\]
\[ s_1=\frac{2x}{x^2+y^2+1},\quad s_2=\frac{1-x^2-y^2}{1+x^2+y^2},\quad s_3=\frac{2y}{x^2+y^2+1}.\]
 \item \[-\frac{dx^2+dy^2}{y^2}=ds_1^2+ds_2^2+ds_3^2.\] 
 \[ s_1=\frac{i}{2y}(1-x^2-y^2),\quad s_2=\frac{ix}{y},\quad s_3=-\frac{1}{2y}(1+x^2+y^2),\]
 \[ x=\frac{s_2}{s_1-is_3},\quad y=\frac{i}{s_1-is_3}.\]
\item\[ -\frac{x^2+y^2}{x^2y^2}(dx^2+dy^2)=ds_1^2+ds_2^2+ds_3^2,\]
\[ s_1=\frac{i}{8}\frac{(x^2+y^2+2)(x^2+y^2-2)}{xy},\ s_2=\frac{i}{2}\frac{(x^2-y^2)}{xy},\ s_3=-\frac18\frac{(x^2+y^2)^2+4}{xy}.\]

\end{enumerate}

\medskip \noindent
  Sphere - sphere
  \begin{enumerate}
   \item \[-\frac{dX^2+dY^2}{Y^2}=-\frac{x^2+y^2}{x^2y^2}(dx^2+dy^2)=ds_1^2+ds_2^2+ds_3^2,\]
   \[ X=\frac{x^2-y^2}{2},\quad Y=xy,\]
   So
   \[ S_2=\frac12(s_2+\frac{1}{s_2}),\quad S_1+iS_3=-\frac{(s_1-is_3)^2}{s_2}.\]
   \item \[ -\frac{dX^2+dY^2}{Y^2}=\left(\frac{4}{(x^2+y^2+1)^2}-\frac{1}{y^2}\right)(dx^2+dy^2).\]
   \[ X=\frac18\frac{x(x^2+y^2+1)}{x^2+y^2},\quad Y=\frac18\frac{y(x^2+y^2+1)}{x^2+y^2}.\]
   \item \[ -\frac{dX^2+dY^2}{Y^2}=\left(\frac{4}{(x^2+y^2+1)^2}-\frac{1}{x^2}\right)(dx^2+dy^2).\]
   \[ X=-\frac{2iy(x^2+y^2+1)}{(x^2+y^2-2y+1)(x^2+y^2+2y+1)}\quad Y=-\frac{2ix(x^2+y^2+1)}{(x^2+y^2-2y+1)(x^2+y^2+2y+1)}.\]
   \[ s_1=-\frac{(x^2+y^2+1)^2+4x^2}{4x(x^2+y^2+1)},\ s_3=i\frac{(x^2+y^2-1)^2-4y^2}{4x(x^2+y^2+1)},\]
   \[ s_2=\frac{iy(x^2+y^2-1)}{x(x^2+y^2+1)}.\]
   \item \[ 4\frac{dX^2+dY^2}{(X^2+Y^2+1)^2}=-\frac{dx^2+dy^2}{y^2},\]
   \[ X=\frac{1}{2}\frac{1-x^2-y^2}{x-iy},\quad Y=\frac{i}{2}\frac{x^2+y^2+1}{x-iy}.\]
   \item \[ -\frac{dX^2+dY^2}{Y^2}=-16\frac{x^2+y^2}{(x^2+y^2-1)^2(x^2+y^2+1)^2}(dx^2+dy^2),\]
   \[ X=\frac{32xy}{(x^2+y^2+2y+1)(x^2+y^2-2y+1)},\quad Y=-\frac{8(x^2+y^2+1)(x^2+y^2-1)}{(x^2+y^2+2y+1)(x^2+y^2-2y+1)}.\]

  \item \[ -\frac{dX^2+dY^2}{Y^2}=\]
  \[-\left(\frac{(x^2+y^2)([x^2+y^2+1]^2-4x^2)([x^2+y^2+1]^2-4y^2)}{x^2y^2(x^2+y^2-1)^2(x^2+y^2+1)^2}\right)
  (dx^2+dy^2),\]
  \[ X=i\frac{(y^2-x^2)([x^2+y^2]^2+1)}{(x^2+y^2)^2},\quad Y=-2i\frac{xy([x^2+y^2]^2-1)}{(x^2+y^2)^2}.\]
  \end{enumerate}

\begin{example}
 Consider the spherical coordinates
 \[ x=\frac{\sin(\theta)\cos(\phi)}{1+\cos(\theta)}, y=\frac{\sin(\theta)\sin(\phi)}{1+\cos(\theta)},\]
 separable for the Laplace system $[1111]$. Under the St\"ackel transform $(1,0,0,0)$ the Laplace equation is mapped to 
 the Helmholtz system $S9$ with metric $ds^2= -\frac{dx^2+dy^2}{x^2}$. Using the flat space - sphere identities $2$ we see that 
 in terms of standard coordinates on the complex 2-sphere  we have $-ds^2=ds_1^2+ds_2^2+ds_3^2$ where 
 \[ s_1=i\cot(\theta)\,\csc(\phi),\ s_2 =i\cot(\phi),\ s_3=-\csc(\theta)\,\csc(\phi),\]
 spherical coordinates on the 2-sphere.  Under the St\"ackel transform $(1,0,0,1)$ the Laplace equation is mapped to 
 the Helmholtz system $S8$ with metric $ds^2= -(\frac{1}{x^2}-\frac{4}{(x^2+y^2+1)^2})(dx^2+dy^2)$. 
 Using the sphere-sphere identities $iii)$ we see that 
 in terms of standard coordinates on the complex 2-sphere  we have $ds^2=ds_1^2+ds_2^2+ds_3^2$ where 
\[ s_1+is_3=-\frac{\sin(\theta)}{\cos(\phi)},\ 
s_1-is_3=\frac12\left(\sin(\theta)\,\cos(\phi)-\frac{1}{\sin(\theta)\,\cos(\phi)}\right),\]
\[ s_2=-i\cos(\theta)\,\tan(\phi),\]
degenerate elliptic coordinates of type 1.
 \end{example}
 
 \begin{example} Consider the Cartesian coordinates $x,y$,
 separable for the Laplace system $[211]$. Under the St\"ackel transform $(0,0,0,1)$ the Laplace equation is mapped to 
 the Helmholtz system $E1$ with metric $ds^2= dx^2+dy^2$, separable in Cartesian coordinates.  Under the St\"ackel transform
 $(1,1,0,0)$ the Laplace equation is mapped to 
 the Helmholtz system $S4$ with metric $ds^2= \left(\frac{1}{x^2}+\frac{1}{y^2}\right)(dx^2+dy^2)$. 
 Using the flat space - sphere identities $3$ we see that 
 in terms of standard coordinates on the complex 2-sphere  we have $-ds^2=ds_1^2+ds_2^2+ds_3^2$ where 
 \[ s_1=\frac{i}{8}\frac{(x^2+y^2+2)(x^2+y^2-2)}{xy},\ s_2=\frac{i}{2}\frac{(x^2-y^2)}{xy},\ s_3=-\frac18\frac{(x^2+y^2)^2+4}{xy}.\]
These are degenerate elliptic coordinates of type 2, separable on the 2-sphere.
\end{example}

The complete identification of Laplace separable systems with  Helmholtz separable systems in constant
curvature spaces is presented in tables \ref{Scoords} and \ref{Ecoords}. The labels $a$-$f$ designate the 6 Laplace 
separable coordinate systems defined in tables \ref{Tab1} and \ref{Tab2}. Note that every separable coordinate system for the zero potential Helmholtz equation $(\partial_x^2+\partial_y^2-E)\Psi=0$ 
on flat space or $(\partial_{s_1}^2+\partial_{s_2}^2+\partial_{s_3}^2-E)\Psi=0$ on the 2-sphere  is also separable for some Helmholtz or Laplace superintegrable system.

\begin{table}[tbhp]
\begin{tabular}{r|*{9}{|c}|}
\makebox[26mm]{}&\bb{S1} &\bb{S2} &\bb{S3} &\bb{S4} &\bb{S5}
                &\bb{S6} &\bb{S7} &\bb{S8} &\bb{S9} \\
\hline\hline
Spherical             &    & $d$ & $a$ & $d$& $d$& $a$ & $a$ &    & $a$ \\
\hline
Horospherical         & $c$ & $c$ & $i$ &    & $c$ &    &    &    &    \\
\hline
Ellipsoidal             & &    & $b$ &    &  & $b$&  & $b$ & $b$\\
\hline
Degenerate elliptic 1 &    & $e$ & $j$ & $e$& $e$ & $j$ & $b$ & $a$ &    \\
\hline
Degenerate elliptic 2 &$g$ &    &    & $c$ & $g$ & $i$ &    &    &    \\
\hline
\end{tabular}
\caption{Separating coordinate systems for superintegrable potentials
on the two-dimensional complex sphere.}
\label{Scoords}
\end{table}

\begin{table}[tbhp]
\begin{tabular}{r|*{10}{|c}|}
\makebox[26mm]{}&\bb{E1}  &\bb{E2}  &\bb{E3}  &\bb{E4}  &\bb{E5}
                &\bb{E6}  &\bb{E7}  &\bb{E8}  &\bb{E9}  &\bb{E10}\\
\hline\hline
Cartesian       & $c$& $c$ & $c$ & $c$ & $c$ &$c$ &    &    &    &    \\
\hline
Polar           &$d$ &    & $d$ &    &    & $d$ &    & $d$ &    &    \\
\hline
Semi-Hyperbolic &    &    &    & $h$ &    &    &    &    &    & $h$ \\
\hline
Hyperbolic      &    &    & $f$ &    &    &    & $d$ & $f$ &    &    \\
\hline
Parabolic       &    & $g$ &    &    & $g$ & $g$ &    &    & $h$ &    \\
\hline
Elliptic        & $e$&    & $e$ &    &    & $e$ & $f$ &    &    &    \\
\hline
\end{tabular}

\vskip1cm

\begin{tabular}{r|*{9}{|c}|}
\makebox[26mm]{}&\bb{E11} &\bb{E12} &\bb{E13} &\bb{E14} 
                &\bb{E16} &\bb{E17} &\bb{E18} &\bb{E19} &\bb{E20} \\
\hline\hline
Cartesian       &    &    &    &        &    &    &    &    &    \\
\hline
Polar           &    &    &    & $d$    &$d$ & $d$ & $d$ &    &    \\
\hline
Semi-Hyperbolic & $c$ &    & $c$ &      &    &    &    &    &    \\
\hline
Hyperbolic      &    & $d$ &    & $f$     &    & $f$ & $f$ & $d$ &    \\
\hline
Parabolic       &    &    & $h$ &        & $c$ &    & $c$ &    & $c$   \\
\hline
Elliptic       &    & $f$ &    &       & $e$ &    & $e$ & $f$ &    \\
\hline
\hline 
\end{tabular}
\caption{Separating coordinate systems for superintegrable potentials
in complex two-dimensional Euclidean space. The exceptional system $E3'$ separates only in the family of Cartesian coordinates c).}
\label{Ecoords}
\end{table}

\section{Limits of separable coordinate systems} 
\begin{center}
\begin{figure}[tbhp]
\vspace{0.in}
\includegraphics[width=12.0cm]{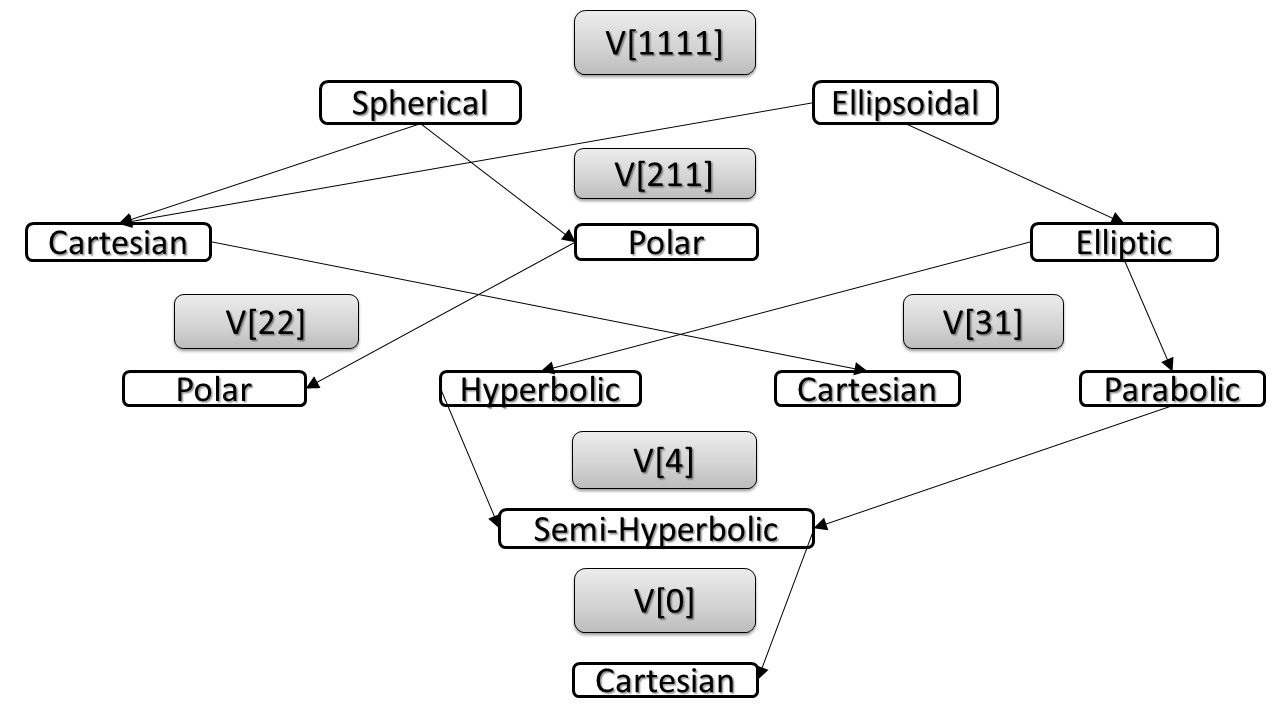}
\caption{B\^ocher contractions of separation equations for nondegenerate  Laplace systems}
\label{Laplacecoords1}
\end{figure}
\end{center}

The separable coordinates for nondegenerate Laplace systems are related by B\^ocher contractions. A B\^ocher contraction
applied to the commuting symmetry operators $H,L=\sum_{j,k}a^{jk}\partial_{x_j,x_j}+\cdots$, characterizing a Laplace separable coordinate system will yield another pair of symmetry operators $H',L'$.
The target symmetries will define a new set of separable coordinates provided the target matrix ${a'}^{jk}({x'}_\ell)$ is diagonalizable. 
The basic results are summarized in Figure \ref{Laplacecoords1}. A separable coordinate system in
a Laplace superintegrable system contracts to a system on a lower level 
provided there is a sequence of directed arrows connecting the source system to the target. This means that there exists a B\^ocher contraction
(or composition of B\^ocher contractions) that takes the symmetry operators for the source system to the symmetry operators for the target. Note that at the top, $V[1111]$, level
we can let $c\to 0$ or $c\to 1$ in the ellipsoidal coordinates to obtain spherical coordinates. In that sense we see that all of the separable coordinates for nondegenerate
Laplace systems are contractions of ellipsoidal coordinates.

\begin{center}
\begin{figure}[tbhp]
\vspace{0.in}
\includegraphics[width=12.0cm]{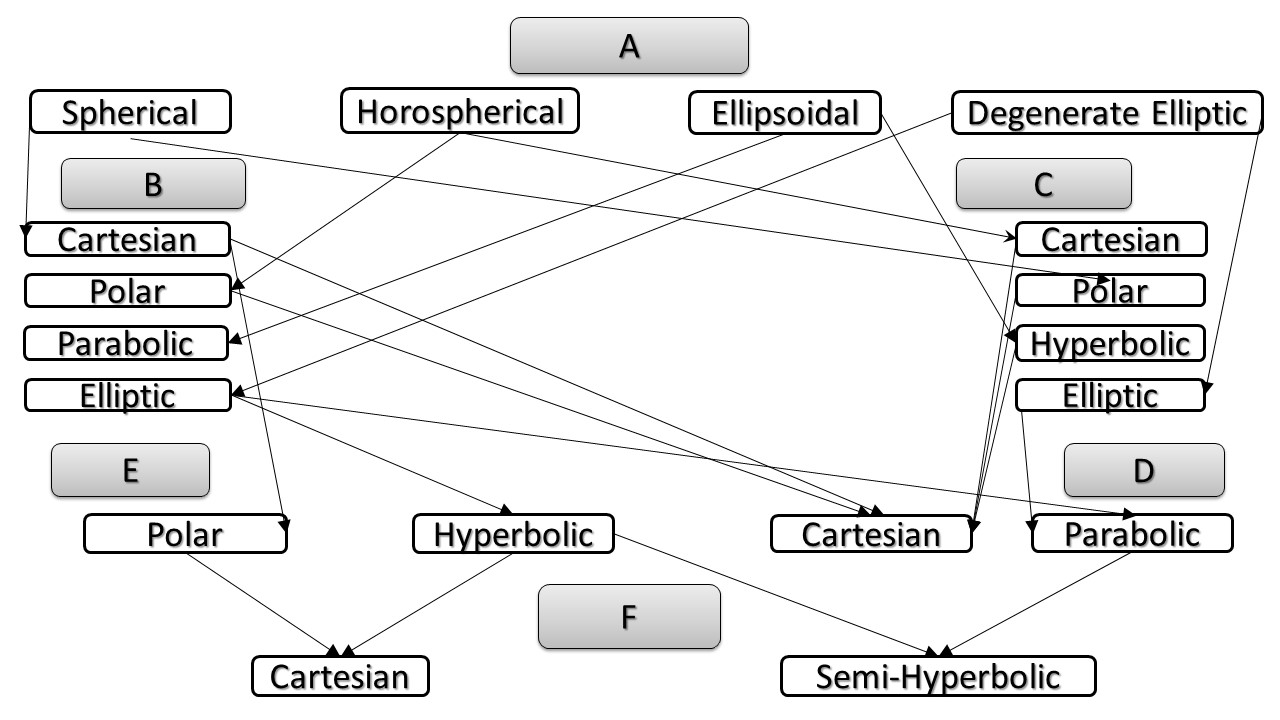}
\caption{Limits of separation equations for degenerate  Laplace systems}
\label{Laplacecoords2}
\end{figure}
\end{center}

The corresponding results for degenerate Laplace systems are  summarized in Figure \ref{Laplacecoords2}. Here there is a difference because of the appearance of 
reduced coordinate systems that can only be obtained by type II limits, \cite{KMR}. These limits are not pure B\^ocher contractions in $x,y$ but also involve limits of the 
separated coordinates. This is accomplished by allowing the parameter $c$ in the generic separable coordinates to be appropriate functions of $\epsilon$, usually 
$c=1+\epsilon^2$ or $c=\frac{1}{\epsilon^2}$.  At the top level $A$ there is a B\^ocher contraction that maps this system into itself. By choosing appropriate functions $c(\epsilon)$ in the 
elliptic 1 coordinates we can obtain the spherical, horospherical and degenerate elliptic coordinates of type 1 in the limit. It follows that all separable coordinates
for degenerate Laplace systems are contraction/limits of elliptic 1 coordinates.

\section{The separation equations}

\begin{center}
\begin{figure}[tbhp]
\includegraphics[width=12.0cm]{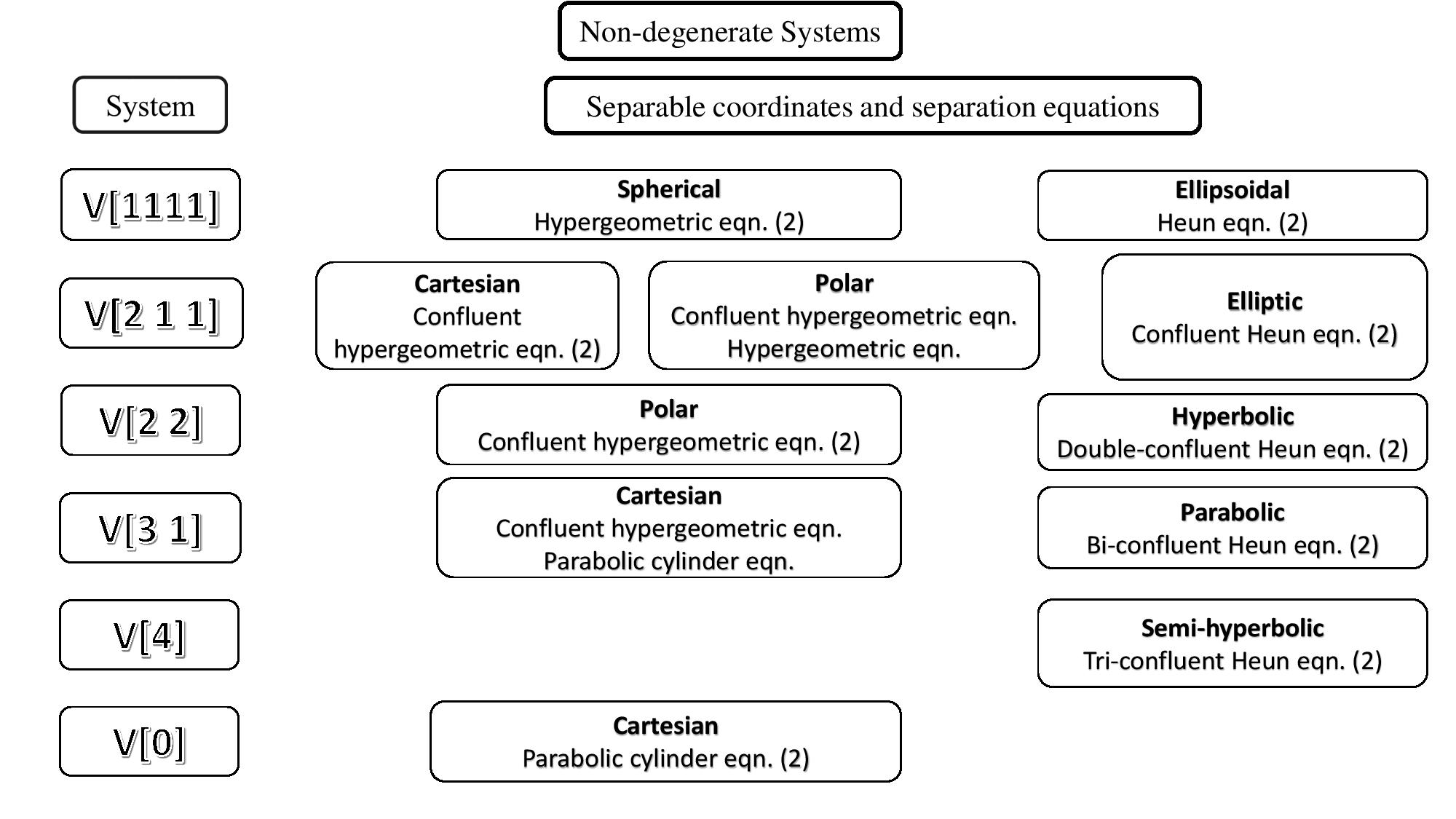}
\caption{Separation equations for nondegenerate Laplace systems }
\label{var0}
\end{figure}
\end{center}

\begin{center}
\begin{figure}[tbhp]
\includegraphics[width=12.0cm]{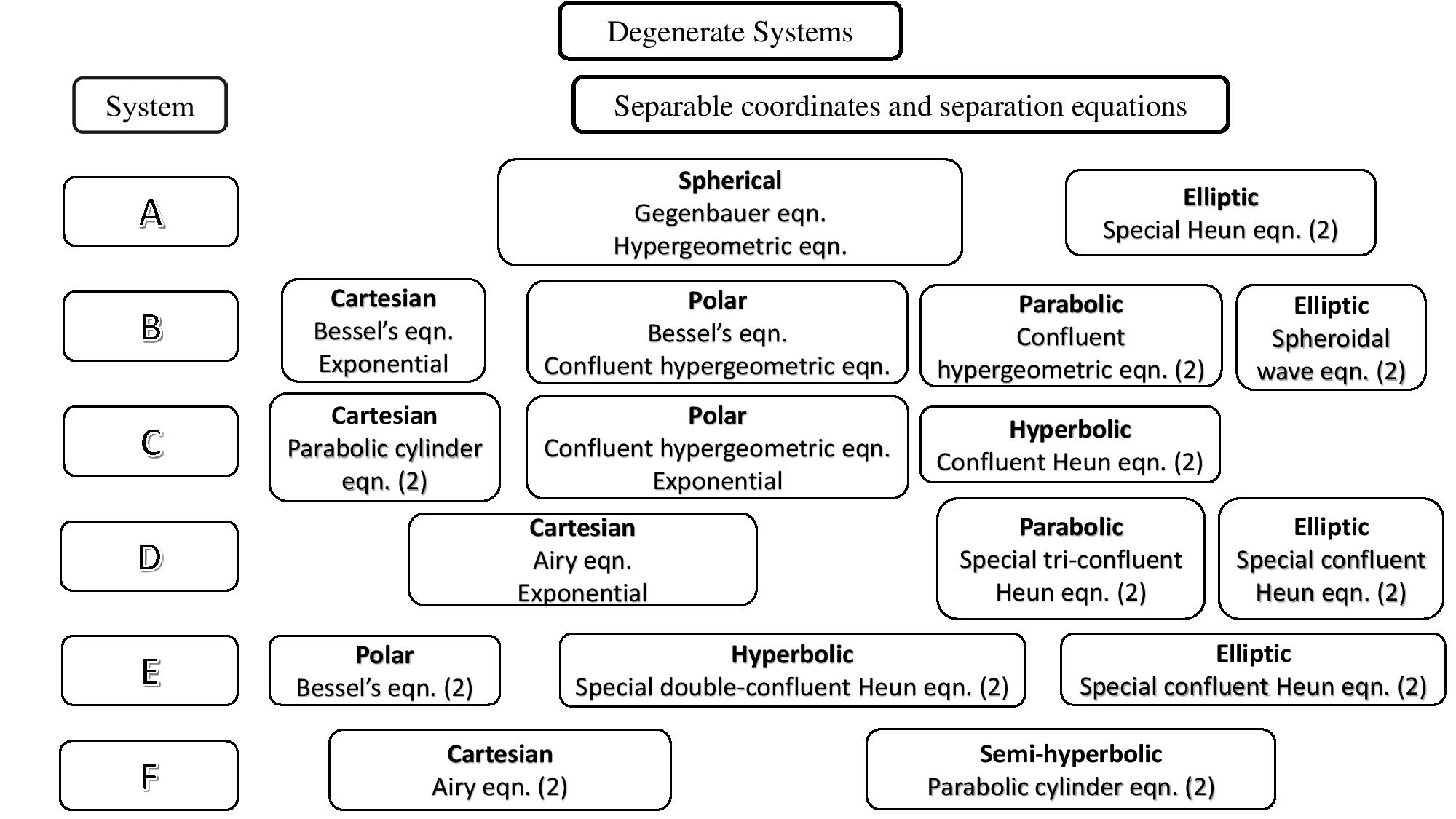}
\caption{Separation equations for degenerate Laplace systems }
\label{var1}
\end{figure}
\end{center}

 Each family of separated solutions for a Laplace or Helmholtz superintegrable system is characterized as the family of eigenfunctions of a 2nd order 
symmetry operator. Each family determines an eigenbasis of separated solutions of the 2D superintegrable system. An eigenbasis of one family 
can be expanded in terms of a eigenbasis for another family and the quadratic structure algebras help to derive the expansion coefficients; 
see e.g., \cite{KM1991}, (with some minor errors), \cite{KMP1996,GVZ,Post2011,IPSW2001,HPSV1999}.  The compete list of separation equations
is given in  Figures \ref{var0} and \ref{var1}. (The notation $(2)$ means that the separation equations for the corresponding coordinates are 
both of the same, except that the separation constant occurs as $c$ in one equation and $-c$ in the other.) The definitions of the 
separation equations are as in \cite{DLMF}.

We first consider the non-degenerate systems in Figure \ref{var0}. Except for the special case of the extended  isotropic oscillator $[0]$, each 
superintegrable system corresponds to exactly one set of coordinates for which the separation equations are of Heun type, i.e., they have 4 regular singular 
points. (They are listed in the right-hand column.)  Moreover, system $[1111]$ corresponds to the general Heun equation, $[211]$ corresponds to the confluent 
Heun equation, $[22]$ to the double-confluent Heun equation, $[31]$ to the bi-confluent Heun equation, and $[4]$ to the triconfluent Heun equation, all in 
full generality with 5 parameters. (The parameters are supplied by the 4 parameters $a_i$ in the non-degenerate potential and the separation constant $c$.) 
Thus any solution of one of these Heun equations determines an eigenfunction of a 2D Laplace or Helmholtz superintegrable system. Similarly the separation 
equations in the left-side columns are all of hypergeometric type, 3 regular singularities. System $[1111]$ corresponds to the general hypergeometric equation, 
system $[211]$ to the hypergeometric and confluent hypergeometric equations, $[22]$ to the confluent hypergeometric equation, $[31]$ to the confluent 
hypergeometric and parabolic cylinder equations, and $[0]$ to the parabolic cylinder equation, all in full generality (including all parameters).

The degenerate systems in Figure \ref{var1} are parameter restrictions of non-degenerate systems for which the restricted system has additional symmetry. 
Except for the special case of the isotropic oscillator $F$, a restriction of $[0]$, each superintegrable system corresponds to at least one set of coordinates 
for which the separation equations are of Heun type. These equations do not have the full number of parameters, but additional symmetry properties. Except for 
the spheroidal wave equation, these special Heun systems do not appear to have individual names but we suggest that they deserve special attention.  
The separation equations in the left-side columns are all of hypergeometric type.  System $A$ corresponds  the  hypergeometric equation and to the Gegenbauer 
equation, system $B$ to the confluent hypergeometric and to Bessel's equation, as well as the equation for exponential functions, $C$ to the confluent 
hypergeometric equation and  the parabolic cylinder equation, as well as the equation for exponential functions, $D$ to the Airy equation, as well as the 
equation for exponential functions, $E$ to Bessel's equation,  and $F$ to the parabolic cylinder equation and the Airy equation, all in full generality.

\section{Hypergeometric type equations}
{\small
\begin{enumerate}
\item Hypergeometric equation:
$z(1-z)\frac{d^2w}{dz^2}+\left(c-(a+b+1)z\right)\frac{dw}{dz}-abw=0$.
	\item Confluent hypergeometric equation:
$z\frac{d^2w}{dz^2}+(b-z)\frac{dw}{dz}-aw=0$.
\item Parabolic cylinder equation:
$\frac{d^2w}{dz^2}+(az^2+bz+c)w=0$.
	\item Gegenbauer equation: 
$ (1-z^2)\frac{d^2w}{dz^2}-2(\mu+1)z\frac{dw}{dz}+(\nu-\mu)(\nu+\mu+1)w=0$.
\item Bessel's equation:
$z^2\frac{d^2w}{dz^2}+z\frac{dw}{dz}+(z^2-\nu^2)w=0$.
\item Airy's equation:
$\frac{d^2w}{dz^2}-zw=0$.		
\end{enumerate}
}
\section{Heun type equations}
{\small
\begin{enumerate}
\item Heun equation:\  $ \alpha+\beta+1=\gamma+\delta+\epsilon$,
\begin{align*} 	&\frac{d^2w}{dz^2}+\left(\frac{\gamma}{z}+\frac{\delta}{z-1}+\frac{\epsilon}{z-a}\right)\frac{dw}{dz}+\frac{\alpha\beta z-q}{z(z-1)(z-a)}w=0,\end{align*}
\item Confluent Heun equation
\begin{align*}&\frac{d^2w}{dz^2}+\left(\frac{\gamma}{z}+\frac{\delta}{z-1}+\epsilon\right)\frac{dw}{dz}+\frac{\alpha z-q}{z(z-1)}w=0.\end{align*}
\item Doubly-confluent Heun equation:
\[ \frac{d^2w}{dz^2}+\left(\frac{\delta}{z^2}+\frac{\gamma}{z}+1\right)\frac{dw}{dz}+\frac{\alpha z-q}{z^2}w=0.\]
\item Biconfluent Heun equation:
$\frac{d^2w}{dz^2}-\left(\frac{\gamma}{z}+\delta+z\right)\frac{dw}{dz}+\frac{\alpha z-q}{z}w=0$.
\item Triconfluent Heun equation:\ 
$ \frac{d^2w}{dz^2}+(\gamma+z)z\frac{dw}{dz}+(\alpha z-q)w=0$.
\item Spheroidal wave equation
\[  	\frac{d}{dz}\left (1-z^2)\frac{dw}{dz}\right)+\left(\lambda+\gamma^2(1-z^2)-\frac{\mu ^2}{1-z^2}\right)w=0.\]

\end{enumerate}
}

\section{Special functions}
 Special functions associated with these systems arise in two distinct ways: 
  \begin{itemize} \item As  separable eigenfunctions of the quantum Hamiltonian.  Second order superintegrable systems are multiseparable. 
   \item As  interbasis expansion coefficients relating distinct separable coordinate eigenbases.  These are often solutions of difference  equations, \cite{MPW2013}
  \end{itemize}
 Most of the special functions in the DLMF appear one of   these ways.    B\"ocher contractions of $S9$ to other 
superintegrable systems  induce limits of the eigenfunctions and  expansion coefficients to  corresponding functions for the contracted superintegrable systems, \cite{KMP2014,KM2014}.
This includes  a reinterpretation of the Askey Scheme 
relating the possible hypergeometric orthogonal polynomials via limits.

\section{Exact and Quasi-exact solvability}
Let $H=\frac{d^2}{dx^2}+V(x)$. We are concerned with the eigenvalue problem $H\Psi=E\Psi$.
The operator $H$ is said to be \textbf{exactly solvable}, \textbf{(ES)} if there exists an infinite flag
  of subspaces of the domain of $H$: ${\cal P}_{N}$, $N=1,2,3,\cdots,$ such that $n_N=\text{dim}{\cal P}_{N}\rightarrow \infty$ as $N\rightarrow \infty$ and
  ${H}\,{\cal P}_{N}\subseteq {\cal P}_{N}\subseteq {\cal P}_{N+1}$ for any $N$. In this case, for each
  subspace ${\cal P}_{N}$ the $n_N$ eigenvalues and eigenfunctions of $H$ can be obtained by pure algebraic means.

  The operator $H$ is called \textbf{quasi-exactly solvable}, \textbf{(QES)} if there exist a single subspace
${\cal P}_{k}$ of dimension $n_k>0$  such that ${H}\,{\cal P}_{k}\subseteq {\cal P}_{k}$. In this case, again we can find $n_k$
eigenvalues and eigenfunctions of $\cal H$ by algebraic means, but we have no information about the remaining eigenvalues and eigenfunctions.
See \cite{TTW:2001, TurbinerQES2,Ushveridze, ExactAndQES, Turbiner:1988}.

There is an intimate connection between these two concepts of solvability, the theory of separation of variables, and 2nd order superintegrability. To illustrate  our treatment we
present the motivating example for QES theory: the  anharmonic oscillator with 6th order potential term:
\be\label{anharmonicosc} H= 
\frac{d^2}{dx^2}+[-\frac{k_1^2}{4\omega^2}+(4n+5)\omega]x^2-k_1 x^4-\omega^2 x^6,\ee
where $n$ is a parameter. Here the eigenvalue equation $H\Psi=\lambda\Psi$ is a Heun equation and explicit series  or integral solutions cannot be found. However, for $n$ a fixed positive integer, there are $n+1$ eigenfunctions
\[\Psi_i=P^{(i)}_n(x^2)e^{-\frac{k_1}{4\omega}x^2-\frac{\omega}{4}x^4},\]
$i=0,1,\cdots,n$ where $P$ is a polynomial of order at most $n$ in  $x^2$, and the eigenvalues and eigenfunctions can be computed by algebraic means. Several theories were developed  to treat this and similar examples, the most prominent being the embedding of the operator $H$ in the enveloping algebra of $s\ell (2,\C)$ or $gl(2,\C)$, \cite{Ushveridze,Turbiner2016}.

However, this example can be understood easily in terms of superintegrability theory, as pointed out in \cite{Letourneau1995}. 
Consider the  singular anisotropic oscillator potential defining the  2nd order superintegrable system  $E2$:
\[
V(x,y)= \frac{1}{ 2}\omega^2(4x^2+y^2) + k_1 x +
\frac{k^2_2 - \frac{1}{ 4}}{ 2y^2}.
\]
The Schr\"odinger equation has the form
\begin{equation}\label{E2Ham}
\left(\frac{\partial^2 }{ \partial x^2} +\frac {\partial^2}{ \partial y^2}
\right)\Psi
+
\left[2E - \omega^2 (4x^2+y^2) - 2k_1 x -
\frac{k^2_2 - \frac14}{ y^2}\right]\Psi = 0.\end{equation}
The Schr\"odinger equation separates in two
systems:  Cartesian and parabolic coordinates. 
Separation of variables  in Cartesian
coordinates leads to  two  independent one-dimensional
Schr\"odinger (separation) equations with Laguerre and Hermite polynomial solutions. These separation equations are ES and one easily computes the energy spectrum 
\[
E = \lambda_1 + \lambda_2 =
\omega [2n + 2 + k_2] - \frac{k_1^2}{8\omega^2},
\qquad n=n_1+n_2 = 0,1,2,...
\]

The  Schr\"odinger equation in parabolic coordinates  is
{\small \[
\frac{1}{ \xi^2 + \eta ^2}
\left(\frac{\partial^2 \Psi }{ \partial \xi^2} +
\frac{\partial^2 \Psi }{ \partial \eta^2}\right)
+ 
\left[2E - \omega^2 (\xi^4 - \xi^2 \eta^2 + \eta^4) - k_1
(\xi^2 - \eta^2) - \frac{k_2^2- \frac14}{ \xi ^2\eta ^2}\right]
\Psi = 0.
\]}
Upon substituting
$
\Psi (\xi, \eta) = X (\xi) Y (\eta)
$
and introducing the parabolic separation constant $\lambda$, we find the  two separation equations:
\be\label{lambda1}
\frac{d^2 X}{d\xi^2}
+ \left(2E \xi^2 - \omega^2\xi^6 - k_1 \xi^4
- \frac{k_2^2 -\frac{1}{4}}{\xi^2}\right)X
=  \lambda X,\ee
\be\label{lambda2}\frac{d^2 Y}{d\eta^2}
+ \left(2E \eta^2 - \omega^2\eta^6 + k_1 \eta^4
- \frac{k_2^2 -\frac{1}{4}}{\eta^2}\right) Y
= -\lambda Y,
\ee
each a form of the bi-confluent Heun equation, in full generality.
Substituting $
E =
\omega [2n + 2 + k_2] - \frac{k_1^2}{8\omega^2}$, and setting $k_2=\frac12,\, \xi=x$ in (\ref{lambda1}) 
we get the QES equation for the anharmonic oscillator with 6th order potential term, where now eigenvalue is the separation constant.
Thus the 1D anharmonic oscillator (\ref{anharmonicosc}) is a special case of the separation equation for $E2$ in parabolic coordinates.

We can verify that the separation equation admits polynomial solutions for these special values of $E$. First we make the change of variable 
$\xi=\sqrt{u}$ in equation (\ref{lambda1}) to obtain
\[ \left[4u\frac{d^2 }{d u^2}+2\frac{d}{du}+\frac{1}{4u}\left(-4\omega^2 u^4-4k_1 u^3+8Eu^2-4k_2^2+1\right)\right]X=\lambda X.\]
Our aim is to modify the operator on the left so that it takes polynomials in $u$ to polynomials. We do this with a gauge transformation:
$X=u^{\frac14+\frac{k_2}{2}}e^{-\frac14 \omega u^2-\frac14 \frac{k_1 u}{\omega}}P(u)$.
This leads to the equation {\small
\[ \left[ 4u\frac{d^2}{du^2}-2(2\omega^2 u^2+k_1u-2k_2\omega-2\omega)\frac{d}{du}\right.\]
\[\left.+\frac{1}{4\omega^2}
 \left(-8k_2\omega^3 u+8 E\omega^2 u-16\omega^3 u+k_1^2 u-4k_1k_2\omega-4k_1\omega\right)\right]P=\lambda P\] }
 where the operator on the left now maps polynomials to polynomials, However in general the oparator doesn't admit invariant subspaces,
 because it increases the degree of a polynomial. The only adjustable parameter here is $E$, so we introduce a new parameter $n$ such that
 $E=\omega(2n+k_2+2)-\frac{k_1^2}{8\omega^2}$.
 Now the preceding equation takes the form
{\small  \be\label{6QES}LP\equiv  \left[ 4u\frac{d^2}{du^2}-2(2\omega^2 u^2+k_1u-2k_2\omega-2\omega)\frac{d}{du}
+\frac{1}{2\omega}
 \left(8n\omega^2 u -3k_1\right)\right]P=\lambda P.\ee}
Note that if $n$  is a nonnegative integer the operator maps the $(n+1)$-dimensional  space of polynomials of order $\le n$
into itself. Thus for each $n$ we can use algebraic methods to determine $n+1$ eigenvalues of $L$ and the corresponding eigenvectors.

In papers \cite{ExactAndQES} a connection was made between the problem of finding QES systems in 1D and 2nd order superintegrability in $n$D. There
it was shown that ``most" generic superintegrable systems in $n$ variables were exactly solvable in some coordinate system, which allowed the 
explicit determination of the discrete bound state energy eigenvalues $E_i$. These systems were all multiseparable. In some cases the 1D separation equations 
were exactly solvable. However, some equations corresponded to QES systems. In these cases the energy eigenvalue $E$ became a parameter in the 1D potential 
and the $1D$ energy was the separation constant $c$. It was shown that the values $E_i$ were exactly those such that the 1D equation $H\Psi=c\Psi$  had QES
polynomial solutions. 

In two recent papers \cite{Turbiner2016, TurbinerHeun}, Turbiner has studied and reported on the classification of QES systems in 1D.
His emphasis is on QES systems that are special cases of the Heun equation and its confluent forms, and exactly solvable systems which are special cases of the 
hypergeometric equation.  We see now that all of these systems correspond to separation equations for the 2D 2nd order superintegrable systems as given in
Figures \ref{var2} and \ref{var3}. Thus all of these solutions determine solutions of the 2D superintegrable systems. 

\subsection{A superintegrable non-Heun example} Second order nD superintegrable systems provide many examples of QES equations for $n>2$ that are not 
Heun equations \cite{ExactAndQES}. We give an example for $n=4$. The flat space superintegrable system is, in Cartesian coordinates $x_j$,
\[ H=\sum_{j=1}^4 \partial_{x_j}^2+\left[\frac14(x^2_2+x^2_3+x^2_4)+x^2_1-2ix_1-3\right]a_5-(ix_1-2)a_4+ \frac{4b_1}{(x_3+ix_4)^2}\]
\[-
\frac{8x_3b_2}{(x_3+ix_4)^3}+\frac{4c_1}{x^2_2}.\]
   We introduce separable coordinates $z_j$, defined by
  \[ x_1=-i(S_1-2),\ x^2_2=4S_4,\]
  \[ (x_3+ix_4)^2=4(S_4-S_3+S_2-S_1+1),\ x^2_3+x^2_4=12-8S_1+4S_2-4S_4,\]
\[ S_1=z_1+z_2+z_3+z_4,\ S_2=z_1z_2+z_1z_3+z_1z_4+z_2z_3+z_2z_4+z_3z_4,\]
\[S_3=z_1z_2z_3 + z_4z_2z_3 + z_1z_2z_4 + z_1z_4z_3,\ S_4=z_1z_2z_3z_4,\]
and  
\[ \sum_{j=1}^4 dx^2_j=\sum_{k=1}^4\Pi_{\ell\ne k}(z_k-z_\ell)\frac{dz^2_k}{ z_k(z_k-1) ^2}.\]
The separation equations are:
\[ \left(\partial_{z_j}^2+\sum_{k=0}^5a_kz_j^k+\frac{b_1}{z_j-1}+\frac{b_2}{(z_j-1)^2}+\frac{c_1}{z_j}\right)Z(z_j)=0,\]
for $j=1,\cdots,4$.  Modulo a gauge transformation of the form $Z(z_j)=R_j(z_j)F_j(z_j)$ with
$R_j=\exp(c_1 z_j^2+c_2z_j+\frac{c_3}{z_j-1})(z_j-1)^{c_4}
z_j^{c_5}$ each of these equations is QES in polynomial variables $z_j$,\cite{ExactAndQES}, but not of Heun type.

\section{Relation of an explicitly solvable QES 1D system to the superintegrable system E2}\label{system31} Some special cases of Heun equations 
reduce to hypergeometric equations,  see the impressive work of Maier,  \cite{Maier}. Moreover, in  recent papers it has been shown that some special 
cases of the Heun equations have explicit solutions that are expressible in terms of derivatives of hypergeometric functions. 
These also yield explicit solutions of 1D QES Schr\"odinger equations, e.g.\cite{exactsolninvsqrt}. We can observe that all such special solutions 
lead to eigenfunctions  of 2D superintegrable systems which also have separable ES hypergeometric eigenfunctions. The quadratic algebras of the 2D systems 
allow us to relate the QES and ES systems. Moreover a knowledge of the possible ES systems for a 2nd order superintegrable system gives important clues 
about the structure of the QES systems

As an example  consider  inverse square root system \cite{exactsolninvsqrt}, written in the form
\begin{equation}\label{invsqrta}  \frac{d^2 f(x)}{dx^2} + (\frac{a}{x^{1/2}}+\frac{b}{x}+\frac{c}{x^{3/2}}-E)f(x).\end{equation}
With $y=\sqrt{x}$ we have
\begin{equation}\label{invsqrtb} y \frac{d^2 f(y)}{dy^2}-\frac{df(y)}{dy} + (4ay^2+4by+4c-4Ey^3)f(y).\end{equation}
Changing notation, we note that the superintegrable system $E2$, (\ref{E2Ham}), in Cartesian coordinates $y_1,y_2$ is 
\begin{equation}\label{E2c} \left(\frac{\partial ^2 }{\partial y_1^2} +  \frac{\partial ^2 }{\partial y_2^2} +  
  (-A(4y_1^2+y_2^2)+By_1+\frac{C}{y_2^2}-E')\right)f(y_1, y_2)=0.\end{equation}
    In these coordinates the separable solutions 
  $f(y_1,y_2)=g_1(y_1)g_2(y_2)$ are eigenfunctions $L_1f=\lambda_1 f$ of the symmetry operator
  $ L_1 =\partial_{y_1^2} + (-4Ay_1^2+By_1)$,                           with separation equations 
  \begin{equation}\label{E2carta} \frac{d^2 g_1}{dy_1^2}+(-4Ay_1^2+By_1+\lambda_1-E')g_1=0,\
    \frac{d^2 g_2}{dy_2^2}+(-Ay_2^2+\frac{C}{y_2^2}-\lambda_1)g_2=0.\end{equation}
   Here, $\lambda_1$ is the separation constant.
   
   In parabolic coordinates $\eta,\xi$ with $y_1=\xi+\eta,y_2=2i\sqrt{\xi\eta}$, (\ref{E2c}), 
   and with a gauge transformation for convenience, $f(\eta,\xi) =(\eta\xi)^{-3/4} f_1(\eta)f_2(\xi)$, the separation equations (\ref{lambda1}),(\ref{lambda2}), reduce to 
  the bi-confluent Heun separation equation
  \begin{equation}\label{E2pa} \eta\frac{d^2 f_1(\eta)}{d\eta^2}-\frac{d f_1(\eta)}{d\eta}+\left(\frac{15}{16\eta}+\frac{C}{4\eta}-E'\eta+B\eta^2-4A\eta^3+\lambda_2\right)
   f_1(\eta)=0,\end{equation}
   for $f_1(\eta)$ with a similar equation for $f_2(\xi)$ (with the separation constant $\lambda_2$ replaced by $-\lambda_2$).
   Now note that, with the restriction to the superintegrable system $E2$ with $C=-15/4$ in the potential, equations (\ref{invsqrtb})
   and (\ref{E2pa}) are the same, provided we make the identifications 
   \begin{equation}\label{ident} y=\eta,\quad 4b=-E',\quad 4a=B,\quad 4c=\lambda_2,\quad E=A.\end{equation}
   Thus this 1D inverse square root potential system corresponds to a special case of the separation equation for the 2D superintegrable system $E2$ in 
   parabolic coordinates. We have already shown that the  separation equation is QES, but we will now show that it has explicit solutions which are a consequence of the exactly solvable $E2$ separation 
   equation (\ref{E2carta}) in Cartesian coordinates.
   
   Making the substitutions (\ref{ident}) where appropriate and setting $y_1=y$, we can solve (\ref{E2carta}) directly to get 
   $g_1(y)=\exp(-y(-a+Ey)/\sqrt{E})\ G(y)$ where $G(y)$ is an arbitrary  linear combination of
      \begin{equation}\label{soln1a}  {}_1F_1\left(\begin{array}{c} \frac{1}{8}(\frac{2E^{3/2}+(4b-\lambda_1)E-a^2}{ E^{3/2}})\\ 
   1/2 \end{array} ; \frac{(2Ey-a)^2}{2E^{3/2}}\right)\end{equation}
   and
   \begin{equation}\label{soln2a}(2Ey-a)\  {}_1F_1\left(\begin{array}{c} \frac{1}{8}(\frac{6E^{3/2}
   +(4b-\lambda_1)E-a^2}{ E^{3/2}})\\ 
   3/2 \end{array} ; \frac{(2Ey-a)^2}{2E^{3/2}}\right).\end{equation}
   Here $G(y)$ is the general solution of the equation $S_1\, G(y)=0$, equivalent to (\ref{E2carta}), where 
   \[ S_1= \frac{d^2}{dy^2}-\frac{2(2Ey-a)}{\sqrt{E}}\frac{d}{dy}-\frac{(2E^{3/2}+4bE-\lambda_1E-a^2)}{E}.\]
   
   The following construction works for $b=-4c^2, \lambda_1=-2\sqrt{E}-32c^2$, which we now assume. Then
   equation (\ref{invsqrtb}) for the inverse square root system is equivalent to $S_2f=0$ where
   \[S_2=y\frac{d^2}{dy^2}-\frac{d}{dy}+4(y^2a-4c^2y+c-Ey^3).\]  
   We define operators $K$ and $Q$ by
   \[ K=\exp(\frac{ay}{\sqrt{E}}-y^2\sqrt{E})\left(\frac{d}{dy}-4y\sqrt{E}+4c+\frac{a}{\sqrt{E}}\right),\]
   \[ Q=\exp(\frac{ay}{\sqrt{E}}-y^2\sqrt{E})\left(y\frac{d}{dy}-4y^2\sqrt{E}+4cy+\frac{ay}{\sqrt{E}}-1\right).\]
   Then it is tedious but straightforward to verify the operator identity
   $S_2K=QS_1$.
   This shows that $K$ maps the solution space of the restricted Cartesian separation equation (\ref{E2carta}) to the solution 
   space of the restricted parabolic separation equation  and provides explicit solutions for the 1D inverse square root potential.
   
   An analogous treatment can be given for the singular Lambert potential 
   \[ V=\frac{V_0}{1+1/W(-e^{-(x+\sigma)/\sigma})},   \quad We^W=x,\]
   considered in \cite{I2016,LB}. Here,  separation in elliptic coordinates for the Laplace system $[211]$ are involved, and 
   the mapping is from the solution space of the restricted Cartesian separation equation (confluent hypergeometric equation) to the solution space
   of the restricted confluent Heun equation.

\section{Conclusions and Outlook}
The theory of 2D 2nd order superintegrable Laplace systems encodes all the information
about 2D Helmholtz or time-independent Schr\"odinger   superintegrable systems in an efficient manner: there is a 1-1 correspondence between Laplace superintegrable systems
and St\"ackel equivalence classes of Helmholtz superintegrable systems. Each of these systems admits a quadratic symmetry algebra  and is multiseparable. The separation equations are identically the same for all Helmholtz systems in an equivalence class, up to a permutation of the energy eigenvalue and the parameters in the potential.
The separation equations comprise all of the various types of hypergeometric and Heun equations in full generality. In particular, they coincide with all of the 1D 
Schr\"odinger exactly solvable (ES) and quasi-exactly solvable (QES) systems related to the Heun operator and its limits.  The separable solutions of these equations are the special functions of mathematical physics. The different systems are related by St\"ackel transforms, by their symmetry algebras
and by B\"ocher contractions of the conformal algebra $so(4,\C)$ to itself, which enables all of these systems to be derived from a single one: the generic potential on the 
complex 2-sphere.  The ES separation equtions are intimately related to the QES equations: They are B\^ocher contractions of QES-type  systems and can be used to determine the values of the parameters for the QES systems and, sometimes to provide a basis of solutions for the QES systems, not just a single solution. This approach facilitates a unified view of special function theory, incorporating hypergeometric and Heun functions in full generality. We have shown that all Heun equations are associated with 2D superintegrable systems and quadratic symmetry algebras, and this added structure can be used to obtain new results for Heun functions. Whereas hypergeometric special functions have been studied intensively, relatively little attention has been paid to Heun functions. The theory of 2nd order superintegrable systems places the two types of separation equations on the same level and interrelates them via symmetry algebras and contractions. 

All of our considerations generalize to 2nd order superintegrable systems in 3D and higher dimensions, but the details are more complicated and much remains to be done. In 3D there is already a  classification of nondegenerate systems and their contractions, \cite{CK2014,CKP}, but not yet of the degenerate cases. The separation equations in $n$D for $n\ge 4$
involve, in general, non-Heun equations. However, the QES theory described in \cite{ExactAndQES} is still applicable. 
\section{Acknowledgment}
This work was partially supported by a grant from the Simons Foundation (\#412351, Willard Miller, Jr)  and by CONACYT grant (\# 250881 to M.A. Escobar ).

\section{Appendix A: Superintegrable systems on $E_{2\C}$ and $S_{2\C}$}
{\small
{\bf Nondegenerate functionally linearly independent  $E_{2\C}$  systems: 
 $H\Psi=(\partial_x^2+\partial_y^2+V)\Psi=E\Psi$.}
\begin{align*}
1)&\,E1:\ V=\alpha(x^2+y^2)+\frac{\beta}{x^2}+\frac{\gamma}{y^2},\ 2) \, E2:
V=\alpha(4x^2+y^2)+\beta x+\frac{\gamma}{ y^2}, \\
3)&\, E3':\ V=\alpha(x^2+y^2)+\beta x+\gamma y,\\
4)&\, E7:\ V=\frac{\alpha(x+iy)}{\sqrt{(x+iy)^2-1}}+\frac{\beta (x-iy)}{\sqrt{(x+iy)^2-1}\ \left(x+iy+\sqrt{(x+iy)^2-1}\ \right)^2}+\gamma (x^2+y^2),\\
5)&\,  E8:\ V=\frac{\alpha (x-iy) }{(x+iy)^3}+\frac{\beta}{(x+iy)^2}+\gamma(x^2+y^2),\
6)\, E9:\ V=\frac{\alpha}{\sqrt{x+iy}}+\beta y+\frac{\gamma (x+2iy)}{\sqrt{x+iy}},\\
7)&\, E10: \ V=\alpha(x-iy)+\beta (x+iy-\frac32(x-iy)^2)+\gamma(x^2+y^2-\frac12(x-iy)^3),\\
8)&\,  E11:\  V=\alpha(x-iy)+\frac{\beta (x-iy)}{\sqrt{x+iy}}+\frac{\gamma }{\sqrt{x+iy}}, \\
9&)\, E16:\ V=\frac{1}{\sqrt{x^2+y^2}}(\alpha+\frac{\beta}{y+\sqrt{x^2+y^2}}+\frac{\gamma}{y-\sqrt{x^2+y^2}}),\\
10)& \,  E17:\ V=\frac{\alpha}{\sqrt{x^2+y^2}}+\frac{\beta}{(x+iy)^2}+\frac{\gamma}{(x+iy)\sqrt{x^2+y^2}},\\
11)&\,  E19: \ V=\frac{\alpha(x+iy)}{\sqrt{(x+iy)^2-4}}+\frac{\beta}{\sqrt{(x-iy)(x+iy+2)}}+\frac{\gamma}{\sqrt{(x-iy)(x+iy-2)}},\\
12)&\,  E20:\ V=\frac{1}{\sqrt{x^2+y^2}}\left(\alpha+\beta \sqrt{x+\sqrt{x^2+y^2}}+\gamma \sqrt{x-\sqrt{x^2+y^2}}\right),
\end{align*}

{\bf Degenerate $E_{2\C}$ systems: 
 $H\Psi=(\partial_x^2+\partial_y^2+V)\Psi=E\Psi$.}
\begin{align*}
 1)&\,E3:\ V=\alpha(x^2+y^2),\ 2)\, E4:\   V=\alpha(x+iy),\ 3)\, E5:\  V=\alpha x, \\
4)&\,E6:\  V=\frac{\alpha}{x^2},\ 5)\, E12:\  V=\frac{\alpha (x+iy)}{\sqrt{(x+iy)^2+c^2}},\ 6)\, E13:\  V=\frac{\alpha}{\sqrt{ x+iy}},\\
7)&\,E14:\   V=\frac{\alpha}{(x+iy)^2},\ \  8)\,E18:\  V=\frac{\alpha}{\sqrt{x^2+y^2}}.
\end{align*}
}

{\small
{\bf Nondegenerate systems on the complex 2-sphere: $H\Psi=(J_{23}^2+J_{13}^2+J_{12}^2+V)\Psi=E\Psi$},\quad 
$J_{k\ell}=s_k\partial_{s_\ell}-s_\ell\partial_{s_k}$, \quad $ s_1^2+s_2^2+s_3^2=1$.
\begin{align*}
1)&\, S1:\  V=\frac{\alpha }{(s_1+is_2)^2}+\frac{\beta s_3}{(s_1+is_2)^2}+\frac{\gamma(1-4s_3^2)}{(s_1+is_2)^4},\\
2)&\, S2: V=\frac{\alpha }{s_3^2}+\frac{\beta }{(s_1+is_2)^2}+\frac{\gamma(s_1-is_2)}{(s_1+is_2)^3},\\
3)&\, S4:\ V=\frac{\alpha}{(s_1+is_2)^2}+\frac{\beta s_3}{\sqrt{s_1^2+s_2^2}}+\frac{\gamma}{(s_1+is_2)\sqrt{s_1^2+s_2^2}},\\
4)&\, S7: \ V=\frac{\alpha s_3}{\sqrt{s_1^2+s_2^2}}+\frac{\beta s_1}{s_2^2\sqrt{s_1^2+s_2^2}}+\frac{\gamma}{s_2^2},\\
5)&\,  S8:\ V=\frac{\alpha s_2}{\sqrt{s_1^2+s_3^2}}+\frac{\beta (s_2+is_1+s_3)}{\sqrt{(s_2+is_1)(s_3+is_1)}}+\frac{\gamma(s_2+is_1-s_3)}{\sqrt{(s_2+is_1)(s_3-is_1)}},\\
6)&\,  S9:\ V=\frac{\alpha}{s_1^2}+\frac{\beta}{s_2^2}+\frac{\gamma}{s_3^2},
\end{align*}
}
{\small
{\bf Degenerate systems on the complex 2-sphere:  $H\Psi=(J_{23}^2+J_{13}^2+J_{12}^2+V)\Psi=E\Psi$},\quad 
$J_{k\ell}=s_k\partial_{s_\ell}-s_\ell\partial_{s_k}$, \quad $ s_1^2+s_2^2+s_3^2=1$.
\begin{align*}
1)&\, S3:\  V=\frac{\alpha }{s_3^2},\ 2)\, S5:\   V=\frac{\alpha }{(s_1+is_2)^2},\
3)\, S6:\   V=\frac{\alpha z}{\sqrt{s_1^2+s_2^2}}.
\end{align*}
}

\section{Appendix B:  $S_{2\C}$ and $E_{2\C}$ separable coordinates}

\subsection{Orthogonal separable coordinates in $S_{2\C}$}
\label{S2C}

As for Euclidean space, coordinates separating the Hamilton-Jacobi equation
on the two-sphere correspond to constants that are 
quadratic in the elements of 
the Lie algebra of its symmetry group $O(3,\C)$.  Coordinates belong to 
the same family if one can be transformed to the other by
a rotation or reflection.  On the complex two-sphere, unlike 
complex Euclidean space,
every quadratic constant, other than a multiple of the Hamiltonian, 
corresponds to a separating coordinate system.  

The separable coordinates on the complex two-sphere and their 
characterizing symmetry operators are:
{\small
\begin{enumerate}
\item Spherical coordinates:
$
s_1=\sin\theta\cos\varphi\,,$ $\quad s_2=\sin\theta \sin\varphi\,,
$
$$
s_3=\cos\theta\,, \qquad  L=J^2_3\,.
$$
\item Horospherical coordinates:
$
s_1=\frac i2\left(v + \frac{u^2-1}v\right)\,, $\quad 
$s_2=\frac12\left(v + \frac{u^2+1}v\right)\,,
$
$$  
s_3=\frac{iu}v\,, \qquad L=(J_1-iJ_2)^2\,.
$$
\item Ellipsoidal coordinates:
$
s_1^2= \frac{(cu-1)(cv-1)}{ 1-c}\,,$\quad $s_2^2= \frac{c(u-1)(v-1)}{ c-1}\,,
$
$$  
s_3^2=cuv\,, \qquad  L=J^2_1+cJ^2_2\,.
$$
\item Degenerate Elliptic coordinates of type 1: 
$
s_1+is_2 = \frac{4cuv}{ (u^2+1)(v^2+1)}\,,$\quad $s_1-is_2 = 
\frac{(u^2v^2+1)(u^2+v^2)}{c uv (u^2+1)(v^2+1)}\,,$
$ 
s_3 =\frac {(u^2-1)(v^2-1)}{ (u^2+1)(v^2+1)}\,, \qquad  L=c^2(J_1+iJ_2)^2-J^2_3\,.
$
\item Degenerate Elliptic coordinates of type 2.
$
s_1+is_2=-iuv\,$,\quad $s_1-is_2=\frac14\frac {(u^2+v^2)^2}{ u^3v^3}\,,
$
$$ 
s_3=\frac i2\frac{u^2-v^2}{uv}\,, \qquad  L=\{J_3,J_1-iJ_2\}\,.
$$
\end{enumerate}
}

\subsection{Orthogonal separable coordinates in $E_{2\C}$ }

Each coordinate system in which the flat space Laplace-Beltrami eigevalue equaton (or the classical Hamilton-Jacobi equation)
is separable on 
$E_{2\C}$ is characterized by a 2nd order symmetry operator.
Coordinate systems that are related by Euclidean group motions
belong to the same family and hence a given family of
coordinates (e.g.\ polar coordinates) 
is associated with an equivalence class of
quadratic elements in the enveloping algebra of $e(2,\C)$.  Two elements
are equivalent if one can be transformed into to other by a combination of
scalar multiplication, addition of multiples of $\partial_x^2+\partial_y^2$ and
Euclidean motions (including reflections). 
The following can be taken as a representative list of
coordinate systems and corresponding constants. (Here $M=x\partial_y-y\partial_x$ and $\{A,B\}=AB+BA$.)
{\small
\begin{enumerate}
\item Cartesian coordinates:
$
x\,,y\,, \qquad     L=\partial^2_x\,.
$

\item Polar coordinates:
$
x_S=r\cos\theta\,,  \quad
y_S=r\sin\theta\,,  \qquad     L=M^2\,.
$
\item Semi-hyperbolic coordinates:
$
x_{SH}=ic(w-u)^2+2ic(w+u)\,, \quad 
y_{SH}=-c(w-u)^2+2c(w+u)\,,
$ 
$
L=\frac{1}{2}\{M,(\partial_x+i\partial_y)\}+c(\partial_x-i\partial_y)^2\,.
$
\item Hyperbolic coordinates:
$
x_H= {r^2+s^2+r^2s^2\over 2crs}$\,,\quad $y_H=i \frac{r^2+s^2-r^2s^2}{2crs}$,\quad 
$L=M^2+c^{-2}(\partial_x+i\partial_y)^2\,.
$
\item Parabolic coordinates:
$
x_P=\xi \eta\,,$
$y_P=\frac12(\xi ^2-\eta ^2)\,,$
\qquad 
$L=\{M,\partial_x\}\,.
$
\item Elliptic  coordinates:
$
x_E=c\sqrt{(u-1)(v-1)}\,,$ $\quad y_E=c\sqrt {-uv}\,,$ \qquad 
$L=M^2+c^{2}\partial^2_x\,.
$
\end{enumerate}
}

\end{document}